\documentclass[a4paper,11pt]{article}
\newcommand{\showchanges}{0}  % Set to 1 for highlights, 0 for clean

\usepackage{xcolor}
\usepackage{soul}
\usepackage{xspace}

\definecolor{addcolor}{rgb}{0.0, 0.5, 0.0}
\definecolor{deletecolor}{rgb}{0.8, 0.0, 0.0}
\definecolor{changecolor}{rgb}{0.0, 0.0, 0.8}

\ifnum\showchanges=1
    \newcommand{\added}[1]{\textcolor{addcolor}{\textbf{#1}}}
    \newcommand{\deleted}[1]{\textcolor{deletecolor}{\st{#1}}}
    \newcommand{\changed}[2]{\textcolor{deletecolor}{\st{#1}} \textcolor{addcolor}{\textbf{#2}}}
    \newcommand{\newsection}[1]{\textcolor{addcolor}{\textbf{[NEW]}} #1}
    \newcommand{\newfig}{\textcolor{addcolor}{\textbf{[NEW FIG]}}\xspace}
    \newcommand{\newtab}{\textcolor{addcolor}{\textbf{[NEW TABLE]}}\xspace}
\else
    \newcommand{\added}[1]{#1}
    \newcommand{\deleted}[1]{}
    \newcommand{\changed}[2]{#2}
    \newcommand{\newsection}[1]{#1}
    \newcommand{\newfig}{}
    \newcommand{\newtab}{}
\fi
\usepackage[english]{babel}
\usepackage{amsmath, amssymb, amsfonts}
\usepackage{graphicx}
\usepackage{float}
\usepackage{authblk} % for multi-author formatting
\usepackage[colorlinks=true, allcolors=blue]{hyperref}
\usepackage{setspace}
\usepackage[margin=2.5cm]{geometry} % <-- Add this line for wider margins
\title{Enhancing Photon Identification with Neural Network Methods}

\author{Yuval Frid}
\author{Liron Barak}
\affil{Tel Aviv University, Tel Aviv-Yafo, Israel}
\date{}

\begin{document}

\maketitle

\begin{abstract}
We investigate photon--pion discrimination in regimes where electromagnetic showers overlap at the scale of calorimeter granularity. Using full detector simulations with fine-grained calorimeter segmentation of approximately $0.025\times0.025$ in $(\eta,\phi)$, we benchmark three approaches: boosted decision trees (BDTs) on shower-shape variables, dense neural networks (DNNs) on the same features, and a ResNet-based convolutional neural network operating directly on calorimeter cell energies. The ResNet significantly outperformed both baseline methods, achieving further gains when augmented with soft scoring and an auxiliary $\Delta R$ regression head. Our results demonstrate that residual convolutional architectures, combined with  \changed{physics-informed loss functions}{physics-motivated training strategies}, can substantially improve photon identification in high-luminosity collider environments in which overlapping electromagnetic showers challenge traditional methods.
\end{abstract}
\section{Introduction}

The Large Hadron Collider (LHC) has enabled precision measurements of Standard Model processes and opened new frontiers in the search for physics beyond the Standard Model. Central to many of these studies is the accurate identification of photons, which provide relatively clean experimental signatures in the high-multiplicity environment of hadron collisions. The clean signature of photons in hadron colliders has made them a powerful probe for precision measurements and searches for new physics~\cite{ATLAS:2012gk, CMS:2012qbp}. Therefore, accurate reconstruction and identification of photons is a crucial task in high-energy physics experiments~\cite{ATLAS:2018egammaPerf}. Photon candidates are identified and reconstructed using calorimeter and tracker information. Due to the nature of photon interactions in the calorimeter -- namely, pair production and bremsstrahlung -- a photon entering the detector produces a cascade of secondary particles, forming an electromagnetic (EM) shower~\cite{Leo:1994Introduction}.

Traditionally, EM showers are characterized using a set of carefully engineered observables known as \emph{shower-shape variables}~\cite{ATLAS:2019PhotonID}. Selection cuts or multivariate classifiers trained on these variables are employed to distinguish \emph{prompt photons}, produced directly in the hard scattering process, from \emph{non-prompt photons} originating from hadronic decays or fragmentation. In both the ATLAS and CMS experiments, boosted decision trees (BDTs) based on these shower-shape variables constitute the standard photon-identification algorithm~\cite{ATLAS:2018egammaPerf,ATLAS:2019PhotonID,CMS}. These approaches have demonstrated excellent performance across most kinematic regimes, but their discriminating power degrades in challenging conditions such as high pile-up (multiple overlapping proton–proton collisions in the same bunch crossing) or when EM showers become spatially overlapping at fine calorimeter granularity. In recent years, machine learning techniques have increasingly been explored for photon and electron identification, aiming to surpass the limitations of handcrafted shower-shape variables. Early studies by the ATLAS and CMS collaborations demonstrated that deep neural networks trained on calorimeter observables can improve discrimination power compared to traditional BDT-based approaches~\cite{ATLAS_CNNPhoton,CMS_DNNPhoton}. Further work has extended these ideas to more expressive architectures such as convolutional neural networks (CNNs) operating directly on calorimeter cell images~\cite{deOliveira2016,Accettura2023}, or hybrid models combining high-level variables with low-level features~\cite{Duarte2018,Komiske2019}. Collectively, these studies highlight the growing potential of deep learning in EM shower classification tasks. However, despite these advances, a persistent challenge in photon identification remains: the discrimination of neutral pion ($\pi^0$) decays that mimic single-photon signatures.

\begin{figure}[H]
\centering
\includegraphics[width=0.5\linewidth]{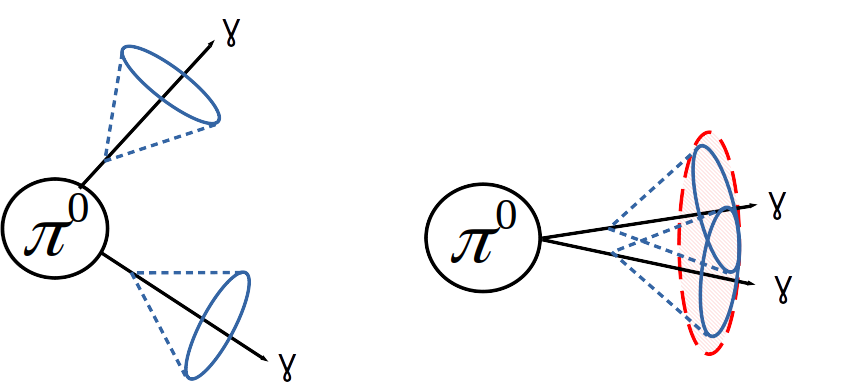}
\caption{\label{fig:Pion_Decay} Illustration of $\pi^0$ decay to two photons for well-separated and highly collimated cases.}
\end{figure}

A notable difficulty arises from the presence of $\pi^0$ mesons, which decay to two photons in over $98\%$ of cases~\cite{PDG2024}. For a $\pi^0$ mass of ${\sim}135\,\text{MeV}$ and photon energies in the GeV range, the opening angle $\theta$ between the two decay photons is typically very small~\cite{ATLAS:2018egammaPerf,PDG2024}. This understanding follows from the invariant mass relation for massless photons:
\begin{equation}
    M_{\pi^0}^2 = 2 E_{\gamma_1} E_{\gamma_2}(1-\cos{\theta}),
\end{equation}
which shows that higher-energy photons correspond to smaller opening angles. As a result, the two decay photons can be highly collimated, sometimes depositing their energy in overlapping regions of the calorimeter, making them difficult to distinguish from single photons (see Fig.~\ref{fig:Pion_Decay})~\cite{ATLAS:2018egammaPerf}.

In this work, we investigate the use of deep neural networks for photon identification in a challenging kinematic regime where standard techniques exhibit reduced performance: photon energies between 10 and 100 GeV, where collimated $\pi^0$ decays are most problematic. Simulated samples are generated with the COCOA-HEP library~\cite{DiBello:2023cocoa}, which couples PYTHIA8~\cite{Sjostrand:2014zea} event generation with full detector simulation in Geant4~\cite{Agostinelli:2002hh}. The calorimeter was configured with three layers in both the electromagnetic and hadronic sections. The EM1 and EM2 layers have nearly symmetric granularity of approximately $0.025 \times 0.025$ in $(\eta, \phi)$, with exact segmentation shown in Table~\ref{table:Cal_Cells}. This near-uniform segmentation provides a suitable two-dimensional shower representation for convolutional neural network studies.

We benchmark three complementary approaches: (i) the standard shower-shape variables with a BDT baseline~\cite{ATLAS:2018egammaPerf}, (ii) a dense neural network trained on the same high-level inputs, and (iii) a ResNet-based convolutional network trained directly on calorimeter images. 
 Additionally, we explore the impact of separating background events according to their diphoton separation $\Delta R$, investigating whether dedicated training for events with $\Delta R < 0.025$ yields further performance gains. While earlier studies have primarily addressed photon–electron or photon–jet separation using deep learning, the present work focuses specifically on the challenging case of $\pi^{0} \to \gamma\gamma$ overlaps at high calorimeter granularity. In addition, we introduce two \changed{physics-informed}{physics-motivated} strategies -- soft scoring and an auxiliary $\Delta R$ regression head—that guide the network toward spatially discriminative features beyond those employed in previous approaches.

\section{Monte Carlo Sample Generation}
\label{sec:MC_Samples}
For this study, a calorimeter simulation with high transverse and longitudinal granularity was utilized to perform photon identification (PID) on detailed shower-shape information. 
While fast-simulation frameworks such as DELPHES~\cite{DELPHES} can be configured to achieve a transverse segmentation comparable to the one used here (down to $\eta \times \phi$ cells of $0.025 \times 0.025$ or finer), they are inherently limited to a single EM and a single hadronic calorimeter layer~\cite{ATL-PHYS-PUB-2017-015,CMS:2017dal}. These layers are modeled as 2D energy maps derived from parameterized detector resolutions, lacking a realistic longitudinal shower development. 
As a result, multi-layer correlations, depth-dependent energy patterns, and realistic shower-shapes cannot be reproduced. Such information is essential for accurate PID, making DELPHES inadequate for the purposes of this study.

Instead, we utilized the \textsc{COCOA-HEP} detector simulation~\cite{DiBello:2023cocoa}, which provides full 3D calorimeter modeling with multiple configurable layers, per-cell energy deposits, and realistic shower evolution. The collision and detector parameters were chosen to be broadly consistent with ATLAS in terms of magnetic field, radiation length, and $\sqrt{s} = 14$~TeV, corresponding to the design energy of the High-Luminosity LHC phase. The calorimeter was configured with three layers in both the EM and hadronic sections, each with uniform $\eta \times \phi$ segmentation as shown in Table~\ref{table:Cal_Cells}.

\added{To maximize the realism of the simulation, electronic noise was implemented via the COCOA framework using layer-dependent Gaussian noise: [13, 34, 41]~MeV for the EM layers and [75, 50, 25]~MeV for the hadronic layers. These values are based on the best imitation of the ATLAS detector environment, ensuring that the simulated energy deposits include the stochastic fluctuations present in experimental data.}

\begin{table}[!h]
\begin{center}
\begin{tabular}{| c  | c|}
 \hline
 COCOA Calorimeter Layer &  $\eta \times \phi$ granularity    \\
 \hline
 EM1  & 0.025 $\times$ 0.0245\\
 \hline
 EM2  & 0.025 $\times$ 0.0245\\
 \hline
 EM3  & 0.050 $\times$ 0.0491\\
 \hline
 HAD1  & 0.100 $\times$ 0.0982\\
 \hline
 HAD2  & 0.100 $\times$ 0.0982\\
 \hline
 HAD3  & 0.200 $\times$ 0.1965\\
 \hline
 \end{tabular}
\caption{Cell granularity in the calorimeter layers in $\eta \times \phi$. The $\phi$ segmentation reflects equal angular divisions of $2\pi$, resulting in values slightly different from the nominal $\eta$ spacing.}
\label{table:Cal_Cells}
\end{center}
\end{table}

\subsection{Signal Samples}
Signal events were generated from five prompt-photon production processes: 
$qg \rightarrow q\gamma$, $qq \rightarrow g\gamma$, $qq \rightarrow \gamma\gamma$, $gg \rightarrow g\gamma$, and $gg \rightarrow \gamma\gamma$~\cite{Sjostrand:2014zea}. 
A minimum hard-scatter transverse momentum (\texttt{pTHatMin}) of 10~GeV was applied, and only photons with $10 < p_T < 100$~GeV were retained. The energy range between 10 and 100~GeV was selected as this is the most challenging regime for photon identification. Above 100~GeV, identification efficiency saturates at a high plateau~\cite{ATLAS:2018egammaPerf}; while below 10~GeV, EM showers are often too small and poorly contained to be reliably separated from hadronic backgrounds, making photon identification practically impossible under realistic detector conditions~\cite{ATLAS:2018egammaPerf,ATLAS:2019PhotonID}. From the full generated dataset, 1.1 million events were selected to produce a flat $p_T$ spectrum, avoiding the over-representation of low-$p_T$ photons inherent in the original steeply falling distribution.  
The dataset was then split into 70\% for training and validation, and 30\% for testing.

\begin{figure}[H]
\centering
\includegraphics[width=0.6\linewidth]{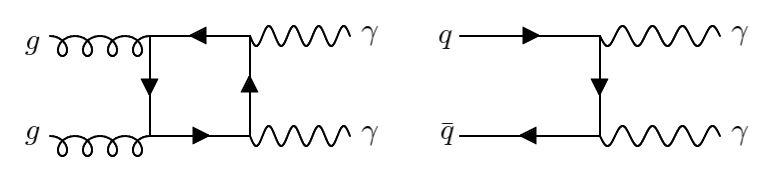}
\caption{\label{fig:Signal_Feynman} The two main channels for direct diphoton production at the LHC.}
\end{figure}

\subsection{Background Samples}
Background events were generated from all QCD processes, requiring at least one $\pi^0$ in the final state with $10 < p_T < 100$~GeV, decaying into two photons~\cite{Sjostrand:2014zea,PDG2024}. The distribution of $\Delta R$ between the two photons is shown in Figure~\ref{fig:soft_score}, with two-thirds of the diphoton pairs having $\Delta R < 0.025$, lower than the detector's resolution. 

A minimum hard-scatter transverse momentum (\texttt{pTHatMin}) of 10~GeV was applied~\cite{Sjostrand:2014zea}. Unlike the signal, no $p_T$ reweighting was performed on the background. This decision preserves the true steeply falling QCD $p_T$ spectrum, which is essential to reflect the dominant low-$p_T$ phase space observed in experimental data.

In total, 4.2 million background events were produced, with 900k used for training and validation, while 3.3 million were reserved for testing. A large test set is essential because, after applying the selection criteria for both signal and background, $\pi^0$ decays occur roughly 1,000 times more frequently than prompt-photon processes. As a result, performance is evaluated in the $\sim$0.1\% fake-rate regime, corresponding with a realistic scenario where the signal-to-background ratio approaches unity after selection.

\begin{figure}[H]
\centering
\includegraphics[width=0.6\linewidth]{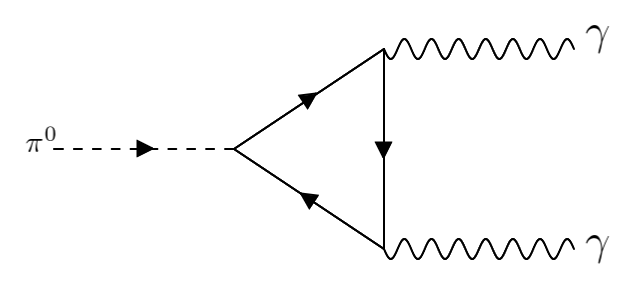}
\caption{\label{fig:Pion_Feynman} $\pi^0$ decay into two photons.}
\end{figure}

The number of events used at each stage is summarized in Table~\ref{table:dataset_summary}.

\begin{table}[H]
\centering
\begin{tabular}{|c|c|c|c|}
\hline
Sample Type & Training & Validation & Test \\
\hline
Signal & 682,470 & 113,670 & 341,370 \\
Background & 700,000 & 200,000 & 3,300,000 \\
\hline
\end{tabular}
\caption{\label{table:dataset_summary} Event counts for each dataset split.}
\end{table}
\section{Methods}
\label{sec:methods}
\subsection{Shower-Shape and BDT Baseline}
Algorithms currently used for object classification in collider experiments are often referred to as Cut-Based algorithms. For each object, a set of shower-shape variables is calculated, and a multi-dimensional cut is optimized to select signal objects at a given fake-positive rate. In practice, modern implementations frequently use a Boosted Decision Tree (BDT) trained on the same variables instead of a literal cut-flow. In this work, we implement the BDT as baseline, using \texttt{XGBoost}~\cite{Chen:2016XGBoost}.
The 20 shower-shape variables used in this work are based on those employed in ATLAS photon identification analyses~\cite{ATLAS:2019PhotonID} and are listed in Table~\ref{table:Cut_Based}. 

\begin{table}[H]
\centering
\begin{tabular}{| c  |    c     |}
 \hline
 Variable Name &  Description   \\
 \hline
 $R_{had}$  &  Ratio of $E_T$ in hadronic layers over $E_T$ in EM layers\\
 $Iso$  &  Ratio of $12\times 12$ $E_T$ sum over $20\times 20$ $E_T$ sum\\
 $F_{EM_X}$  &  Ratio of $E_T$ sum in $EM_X$ over sum in all layers\\
 $R_\eta$  & Ratio of $3\times 7$ $E_T$ sum over $7\times 7$ $E_T$ sum\\
 $R_\phi$  & Ratio of $3\times 3$ $E_T$ sum over $3\times 7$ $E_T$ sum \\
 $W_{\eta2}$  &  Second moment of $\eta$ in $3\times 5$ cells: $\sqrt{(\Sigma E_i \eta_i^2)/(\Sigma E_i)-((\Sigma E_i \eta_i)/(\Sigma E_i))^2}$\\
 $E_{ratio}$  &  Ratio between difference of two maximum cells over their sum \\
 $E_{dR}$  &  $\Delta R$ distance between two maximum cells \\
 \hline
\end{tabular}
\caption{\label{table:Cut_Based}Shower-shape variables for Cut-Based and BDT algorithms. All \(A \times B\) rectangles are in $\eta \times \phi$ space. A single \(1\times 1\) cell corresponds to $0.025\times 0.0245$ in $(\eta, \phi)$.}
\end{table}
The first two variables are per sample, while the last six are per EM layer, resulting in 20 variables in total. 

\subsection{Dense Neural Network Baseline}
To provide a direct comparison with the BDT, we also implement a Dense Neural Network (DNN) trained on the same twenty shower-shape variables. This baseline, implemented in \texttt{PyTorch}~\cite{Paszke:2019PyTorch}, is designed to have a comparable number of trainable parameters to the BDT, ensuring that differences in performance are due to model architecture rather than capacity. The DNN consists of several fully connected layers with ReLU activations and dropout for regularization, as shown in Figure \ref{fig:DNN}.

\subsection{CNN/ResNet Calorimeter-Based Model}
Our primary approach uses calorimeter cell information directly, without engineered features.  
Due to the different granularities of the EM and Had calorimeters (Table~\ref{table:Cal_Cells}), they are treated as two separate input branches:
\begin{itemize}
    \item \textbf{EM branch:} Three layers, with the last one upsampled so that the coarsest layer matches the resolution of the others. An ROI of $0.3 \times 0.295$ in $(\eta,\phi)$ was selected, resulting in a $3\times 12\times 12$ tensor.
    \item \textbf{Had branch:} Similarly upsampled, producing a $3\times 3\times 3$ tensor.
\end{itemize}
Each branch is processed through a sequence of residual convolutional blocks (\textit{ResNet blocks}), allowing the network to learn deep spatial features without vanishing gradients.  
The outputs of the EM and Had branches are flattened, concatenated, and passed through fully connected layers for classification, as shown in Figure \ref{fig:ResNet}.
This two-prong ResNet-based architecture, implemented in \texttt{PyTorch}~\cite{Paszke:2019PyTorch} and using residual blocks~\cite{He:2015ResNet}, is designed to exploit spatial correlations in the calorimeter data, which are inaccessible to cut-based or shower-shape approaches.

\subsection{Additional Training Strategies}
\changed{Beyond the baseline ResNet architecture, we explored two complementary strategies aimed at improving performance in challenging background regimes.}{We explore two physics-motivated training strategies designed to address specific challenges in the overlapping shower regime.}

\begin{figure}[H]
\centering
\includegraphics[width=1\textwidth]{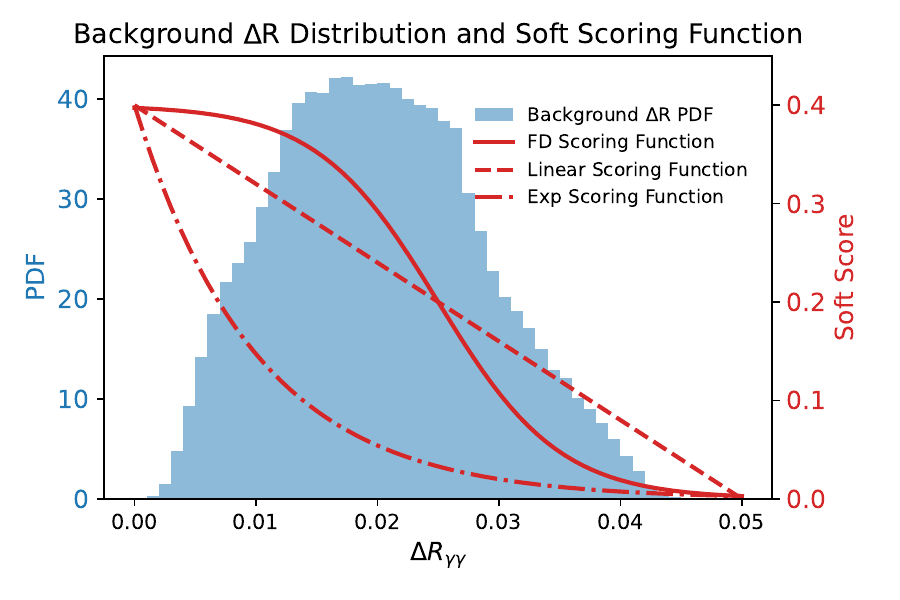}
\caption{Distribution of $\Delta R$ between photon pairs from $\pi^0$ decay (histogram, left axis) and \added{three} soft-scoring target functions (right axis)\deleted{the corresponding soft-scoring target function (red curve, right axis)}. \added{We evaluate Fermi-Dirac (FD), linear, and exponential functional forms, each shown for a representative maximum score amplitude.} Background events \changed{ware}{are} assigned continuous labels between 0 and \added{the maximum amplitude} \deleted{0.4} using \deleted{a Fermi-Dirac-inspired function}\added{these functions}, reducing the penalty for near-threshold cases while maintaining hard labels for well-separated pairs.}
\label{fig:soft_score}
\end{figure}

\paragraph{Soft scoring scheme} 
Background events with photon pairs at very small angular separation ($\Delta R < 0.025$) are particularly difficult to reject due to their high similarity to true photon showers.  
To \changed{encourage the network to better separate these cases}{address label ambiguity in this regime}, we implemented a \emph{soft scoring} scheme (also known as label smoothing)~\cite{Szegedy:2016SoftLabels,müller:smoothing} in which such background events are assigned continuous target labels between~0 and~1 rather than a hard zero, as shown in Figure \ref{fig:soft_score}.  
This smooth target transition reduces the loss penalty for near-misses and allows the model to focus its capacity on improving discrimination where it is most needed. \deleted{Following multiple tests, the Fermi-Dirac (FD) function in the range [0,0.4] achieved the best results.}
\added{We systematically evaluate three functional forms—Fermi-Dirac (FD), linear, and exponential—across a range of maximum score amplitudes (0.1–0.6), with detailed results presented in Section~\ref{subsec:soft_optimization}.}

% Optional figure placeholders

\paragraph{Auxiliary $\Delta R$ head}
\label{subsec:dr_dependence}
We further added an auxiliary output branch that predicts the $\Delta R$ of background photon pairs~\cite{Caruana:1997Multitask}, as shown in Figure \ref{fig:ResNet_dR}.  
This branch shares the same convolutional backbone as the main classifier but has its own regression head, contributing an auxiliary loss term during training.  
The predicted $\Delta R$ does not contribute to the final classification score\deleted{, as it is highly correlated with the output score}. Rather, it serves as an \changed{additional supervision signal to guide the feature extractor towards representations sensitive to small-scale spatial differences}{regularization mechanism, encouraging the shared feature extractor to learn representations sensitive to spatial shower structure}.  
\added{The auxiliary loss weight $\alpha$ controls the relative contribution of classification (BCE) versus regression (MSE) objectives, with systematic evaluation presented in Section~\ref{subsec:dr_optimization}.}
This multi-task setup proved effective in regularizing the network and improving robustness, especially for closely spaced photon pairs.

\subsection{Experimental Protocol}
\label{subsec:experimental_protocol}

\added{To ensure fair comparison across all experiments, we employ controlled initialization and data ordering. All models in the ablation studies (Sections~\ref{subsec:soft_optimization}, \ref{subsec:dr_optimization}, and \ref{subsec:combined}) are initialized from identical random weights (seed=42) and trained on data presented in the same order across epochs. This eliminates initialization variance and ensures that performance differences reflect architectural choices or training strategies rather than stochastic effects. Each configuration is trained for 100 epochs using the Adam optimizer with learning rate $10^{-3}$ and batch size 1000.}
\section{Results}
\label{sec:results}

\added{This section presents comprehensive performance evaluation of the proposed methods, including systematic hyperparameter optimization, ablation studies, and robustness tests under detector systematics.}

We first compare baseline approaches to quantify the gain from using full calorimeter information over engineered shower-shape variables (Section~\ref{subsec:baseline}). 
\added{We then systematically optimize two physics-motivated training strategies: soft scoring to address label ambiguity (Section~\ref{subsec:soft_optimization}) and auxiliary $\Delta R$ supervision for feature regularization (Section~\ref{subsec:dr_optimization}).}
\added{Section~\ref{subsec:combined} evaluates the combination of these techniques, revealing interference effects when using independently-optimized hyperparameters.}
Finally, \added{Section~\ref{subsec:robustness} assesses robustness under realistic detector systematics including energy scale variations and increased noise conditions.}

% ============================================================================
% SUBSECTION 4.1: BASELINE COMPARISON
% ============================================================================
\subsection{Baseline Comparison: BDT, DNN, and ResNet}
\label{subsec:baseline}

\added{This subsection establishes baseline performance for three approaches: shower-shape variables with a BDT, a dense neural network (DNN) using the same inputs, and a ResNet operating directly on calorimeter images. All performance curves include binomial uncertainty bands.}

\begin{figure}[htbp]
\label{fig:baseline_performance}
\centering
\begin{tabular}{cc}
\includegraphics[width=0.48\textwidth]{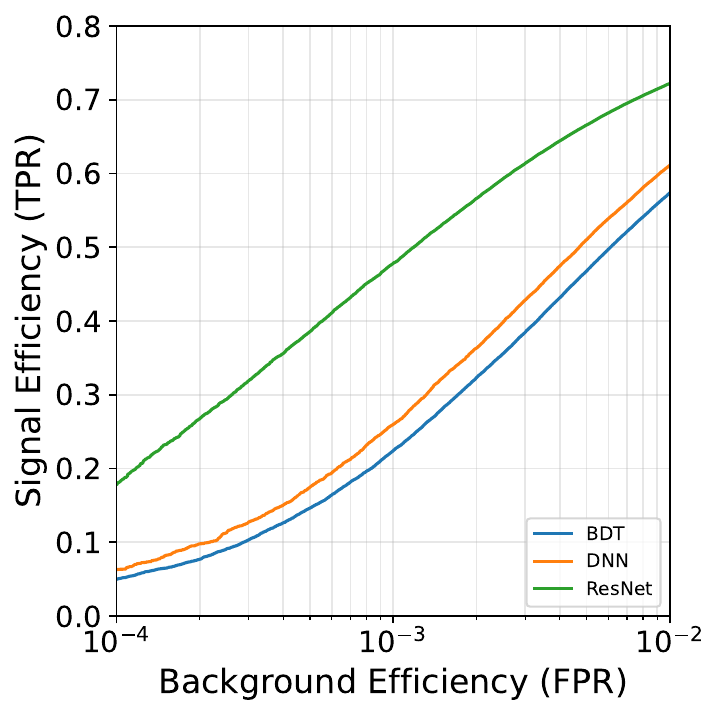} &
\includegraphics[width=0.48\textwidth]{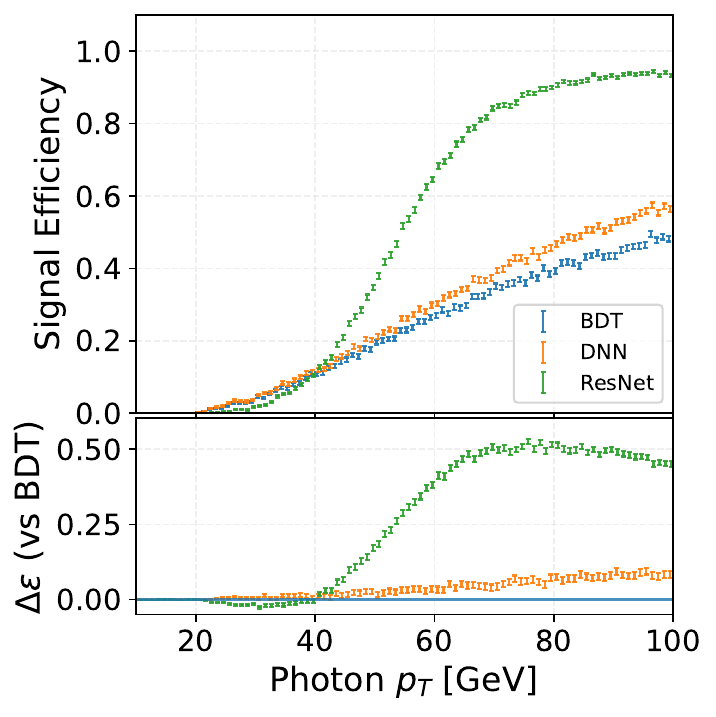} \\
(a) & (b)
\end{tabular}
\end{figure}

Figure~\ref{fig:baseline_performance} presents both ROC and turn-on curves for the baseline methods. 
Panel (a) shows ROC curves for the three baseline models, focusing on the background efficiency range $10^{-4} \leq \epsilon_{\mathrm{bkg}} \leq 10^{-2}$, corresponding to the high-purity regime relevant for photon identification at the LHC. 
The ResNet significantly outperforms both the BDT and the DNN across the entire range, achieving higher signal efficiency at fixed background efficiency.

Panel (b) shows turn-on curves illustrating signal efficiency as a function of true $p_T$ at a 0.1\% false positive rate. 
\added{Error bars represent binomial statistical uncertainties.}
The ResNet maintains a steeper turn-on and higher plateau efficiency than either the BDT or DNN, particularly at low $p_T$, where shower-shape variables alone are less discriminating.

\added{Table~\ref{tab:baseline_performance} quantifies these differences using the $E_{90}$ metric (photon energy at which 90\% signal efficiency is reached), the AUC metric (area under the TOC) and computational benchmarks.}

\begin{table}[htbp]
\newtab
\centering
\caption{\added{Baseline performance and computational metrics. $E_{90}$ is the energy for 90\% signal efficiency at 0.1\% FPR. AUC is the integrated turn-on efficiency (10-100 GeV) as percentage. Inference times measured on NVIDIA GeForce RTX 2070 Super (DNN, ResNet) or CPU (BDT). Training conducted on 1.4M samples over 100 epochs.}}
\label{tab:baseline_performance}
\begin{tabular}{lcccccc}
\hline
\textbf{Method} & \textbf{$E_{90}$} & \textbf{AUC} & \textbf{Inference} & \textbf{Training} & \textbf{Parameters} & \textbf{Device} \\
 & [GeV] & [\%] & [$\mu$s/event] & [hours] & & \\
\hline
BDT    & --- & 21.89 & 0.14 & 0.3 & ---  & CPU \\
DNN    & --- & 25.35 & 0.54 & 0.2 & 49k  & GPU \\
ResNet & 79.3 & 46.79 & 30.1 & 0.9 & 62k  & GPU \\
\hline
\end{tabular}
\end{table}

\added{The ResNet achieves $E_{90} = 79.3$ GeV, representing substantial improvement over both baselines. This demonstrates the benefit of learning directly from spatial calorimeter patterns rather than engineered shower-shape variables.}

\added{Computational benchmarks show the ResNet requires 30.1 $\mu$s/event on GPU, approximately 200× slower than the BDT but still suitable for high-throughput offline analysis. The DNN provides an intermediate option at 0.54 $\mu$s/event. All methods achieve processing rates exceeding 1000 events/second.}        

% ============================================================================
% SUBSECTION 4.2: SOFT SCORING OPTIMIZATION
% ============================================================================
% ============================================================================
% SECTION 4.2: SOFT SCORING OPTIMIZATION (CORRECTED VERSION)
% ============================================================================

\subsection{\newsection{Soft Scoring Optimization}}
\label{subsec:soft_optimization}

\added{To address label ambiguity in the highly collimated background regime ($\Delta R < 0.025$), we systematically evaluate soft-scoring strategies. As described in Section~\ref{subsec:dr_dependence}, soft scoring assigns continuous target labels to background events based on their $\Delta R$, with events at very small separations receiving intermediate scores rather than binary classification. This approach explicitly models the physical ambiguity inherent in distinguishing highly collimated two-photon $\pi^0$ decays from single photons.}

\added{We evaluate three functional forms for the soft-scoring transformation—Fermi-Dirac (FD), linear, and exponential—each tested across six maximum score amplitudes ($s_{\mathrm{max}} = 0.1, 0.2, \ldots, 0.6$), yielding 18 total configurations. The amplitude parameter controls how close to a perfect signal score (1.0) the most collimated background events are allowed to approach. All experiments use identical weight initialization (Section~\ref{subsec:experimental_protocol}) to ensure fair comparison.}

\added{Table~\ref{tab:soft_scoring} presents comprehensive results for all configurations, organized by functional form. For each function, we report $E_{90}$ (lower is better), integrated turn-on AUC (higher is better), and changes relative to the baseline ResNet ($E_{90} = 79.3$ GeV, AUC = 45.5\%). The best result for each metric is marked in \textbf{bold}.}

\begin{table}[htbp]
\newtab
\centering
\caption{\added{Soft scoring systematic evaluation across three functional forms and six maximum score amplitudes. $E_{90}$ is the photon energy at 90\% signal efficiency (0.1\% FPR). AUC is the integrated turn-on efficiency (10-100 GeV). Best result per metric marked in \textbf{bold}.}}
\label{tab:soft_scoring}
\small
\begin{tabular}{lccccccc}
\hline
\textbf{Metric} & \textbf{Base} & \textbf{0.1} & \textbf{0.2} & \textbf{0.3} & \textbf{0.4} & \textbf{0.5} & \textbf{0.6} \\
\hline
\multicolumn{8}{c}{\textbf{Fermi-Dirac Function}} \\
\hline
$E_{90}$ [GeV]  & 79.3 & 79.8 & 84.0 & 79.1 & 76.1 & 75.7 & 75.2 \\
$\Delta E_{90}$ [GeV] & --- & +0.5 & +4.7 & -0.2 & -3.2 & -3.6 & -4.1 \\
AUC [\%]       & 45.5 & 46.4 & 43.2 & 46.1 & 49.2 & 48.6 & 49.2 \\
$\Delta$AUC [\%]   & --- & +0.9 & -2.3 & +0.6 & +3.7 & +3.1 & +3.7 \\
\hline
\multicolumn{8}{c}{\textbf{Linear Function}} \\
\hline
$E_{90}$ [GeV]  & 79.3 & 77.6 & 90.0 & --- & 75.7 & \textbf{74.0} & 74.2 \\
$\Delta E_{90}$ [GeV] & --- & -1.7 & +10.7 & --- & -3.6 & -5.3 & -5.1 \\
AUC [\%]       & 45.5 & 48.1 & 43.2 & 40.2 & 49.1 & \textbf{49.3} & 48.3 \\
$\Delta$AUC [\%]   & --- & +2.6 & -2.3 & -5.3 & +3.6 & +3.8 & +2.8 \\
\hline
\multicolumn{8}{c}{\textbf{Exponential Function}} \\
\hline
$E_{90}$ [GeV]  & 79.3 & 80.1 & 84.7 & 77.5 & 76.0 & 81.3 & 76.5 \\
$\Delta E_{90}$ [GeV] & --- & +0.8 & +5.4 & -1.8 & -3.3 & +2.0 & -2.8 \\
AUC [\%]       & 45.5 & 45.5 & 44.5 & 47.9 & 48.5 & 46.4 & 48.6 \\
$\Delta$AUC [\%]   & --- & +0.0 & -1.0 & +2.4 & +3.0 & +0.9 & +3.1 \\
\hline
\end{tabular}
\end{table}

\added{The results reveal substantial variation in performance across the amplitude range. Within each functional form, some configurations provide improvements over baseline while others degrade performance. For the Fermi-Dirac function, amplitudes 0.4--0.6 yield consistent improvements ($\Delta E_{90} = -3$ to $-4$ GeV), while amplitude 0.2 degrades performance ($\Delta E_{90} = +4.7$ GeV). The linear function shows the strongest improvements at amplitudes 0.4--0.6 ($\Delta E_{90} = -3.6$ to $-5.3$ GeV), but exhibits severe degradation at 0.2 ($\Delta E_{90} = +10.7$ GeV) and fails to reach 90\% at 0.3. The exponential function produces more modest improvements, with amplitudes 0.3--0.4 and 0.6 yielding 2--3 GeV gains.}

\added{The optimal amplitude varies by functional form: Fermi-Dirac peaks at 0.6 ($\Delta E_{90} = -4.1$ GeV), linear at 0.5 ($\Delta E_{90} = -5.3$ GeV), and exponential at 0.4 ($\Delta E_{90} = -3.3$ GeV). The best overall configuration is \textbf{linear 0.5}, achieving $E_{90} = 74.0$ GeV (6.7\% improvement) and AUC = 49.3\% (+3.8\% absolute). This configuration provides both strong threshold performance ($E_{90}$) and robust integrated efficiency (AUC), indicating enhanced discrimination across the full $p_T$ spectrum.}

\begin{figure}[htbp]
\newfig
\centering
\begin{tabular}{cc}
\includegraphics[width=0.48\textwidth]{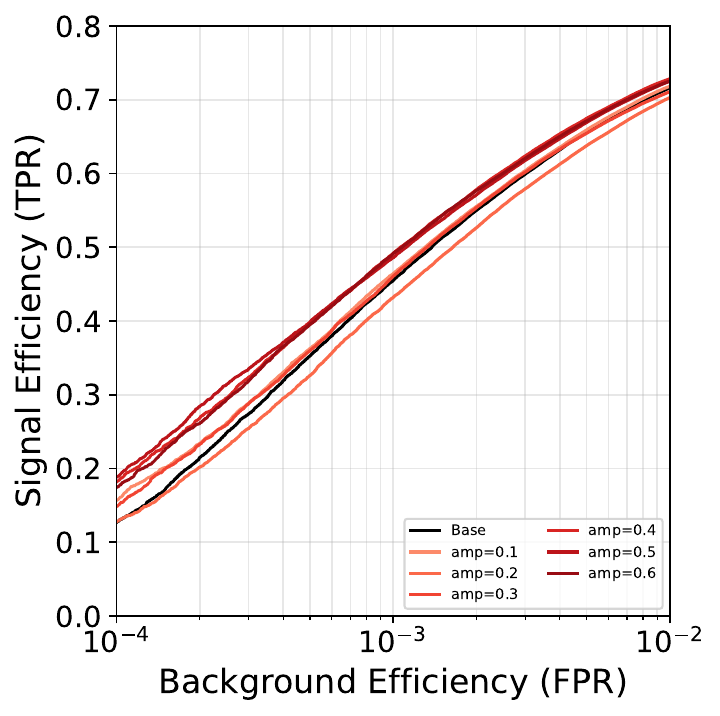} &
\includegraphics[width=0.48\textwidth]{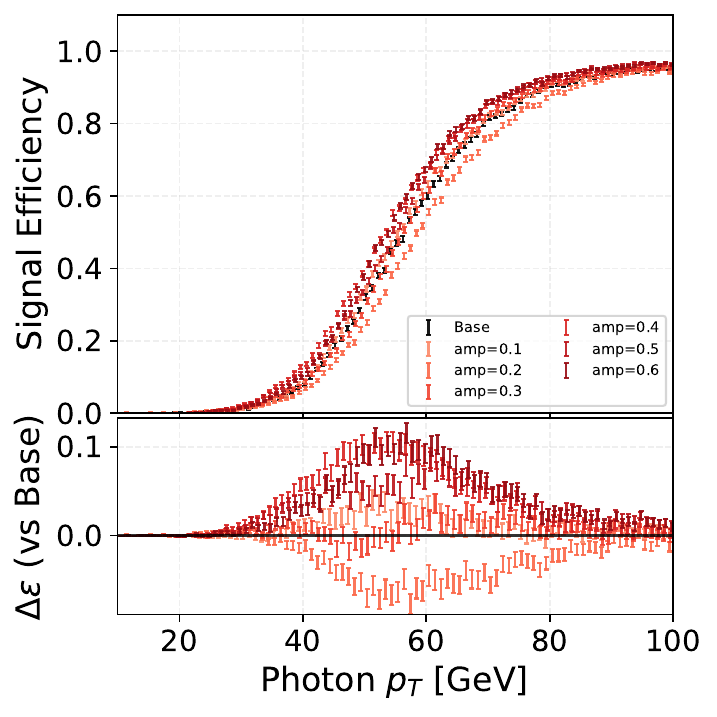} \\
(a) & (b)
\end{tabular}
\caption{\added{Soft scoring with Fermi-Dirac function.} (a) ROC curves showing baseline (black) and six amplitude variations (orange gradient). (b) Turn-on curves at 0.1\% false positive rate with error bars and residuals vs.\ baseline. Amplitudes 0.4--0.6 improve performance ($\Delta E_{90} = -3$ to $-4$ GeV), while 0.2 degrades it ($\Delta E_{90} = +4.7$ GeV).}
\label{fig:soft_fd}
\end{figure}

\begin{figure}[htbp]
\newfig
\centering
\begin{tabular}{cc}
\includegraphics[width=0.48\textwidth]{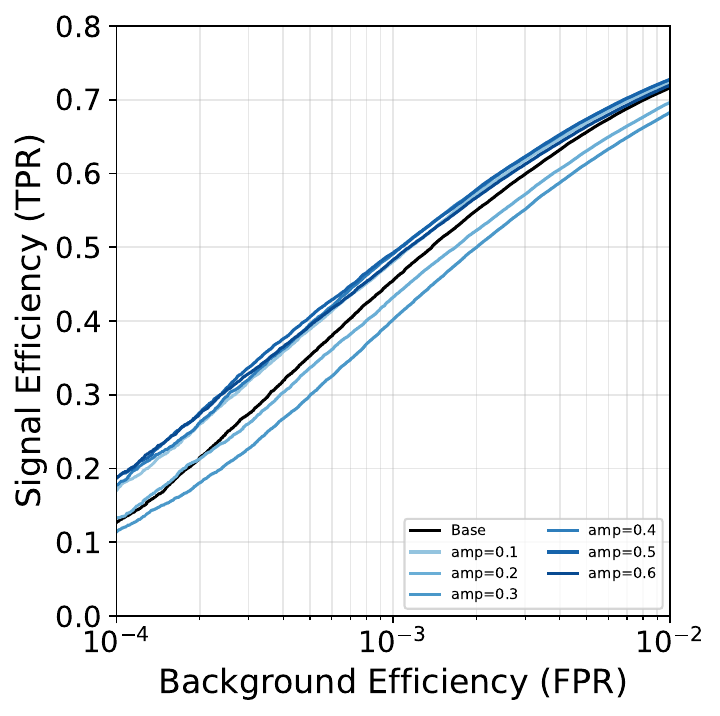} &
\includegraphics[width=0.48\textwidth]{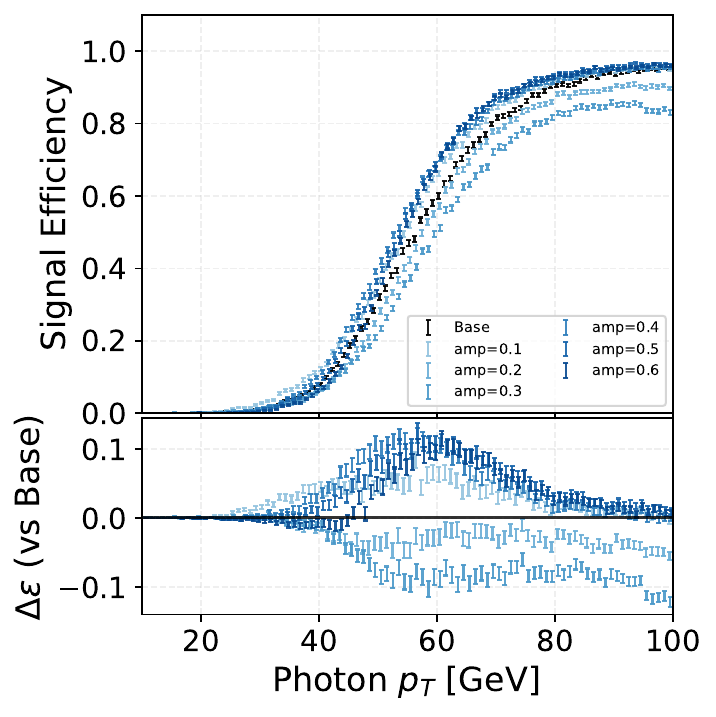} \\
(a) & (b)
\end{tabular}
\caption{\added{Soft scoring with linear function.} (a) ROC curves showing baseline (black) and six amplitude variations (blue gradient). (b) Turn-on curves at 0.1\% false positive rate with error bars and residuals vs.\ baseline. Amplitude 0.5 achieves optimal performance ($E_{90} = 74.0$ GeV, $\Delta E_{90} = -5.3$ GeV, AUC = 49.3\%), while 0.2 severely degrades it ($\Delta E_{90} = +10.7$ GeV) and 0.3 fails to reach 90\% at all.}
\label{fig:soft_linear}
\end{figure}

\begin{figure}[htbp]
\newfig
\centering
\begin{tabular}{cc}
\includegraphics[width=0.48\textwidth]{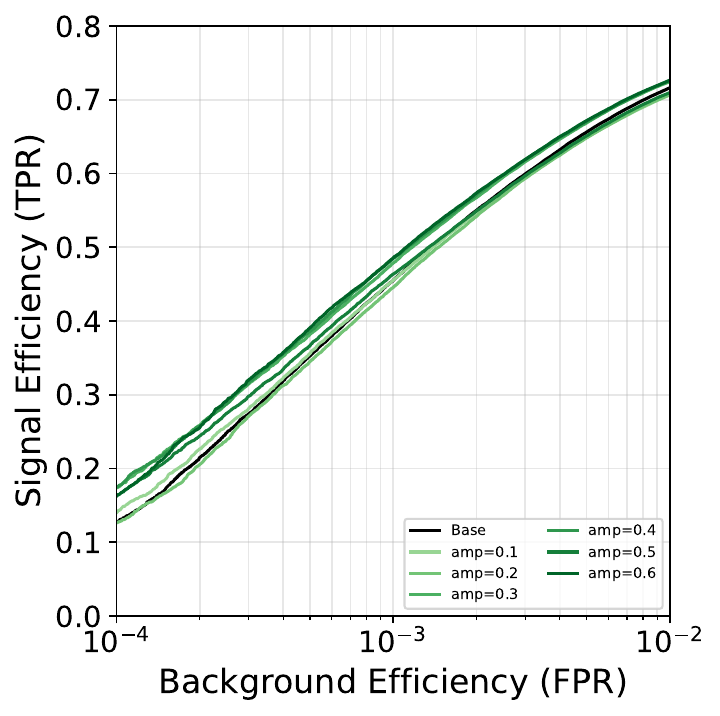} &
\includegraphics[width=0.48\textwidth]{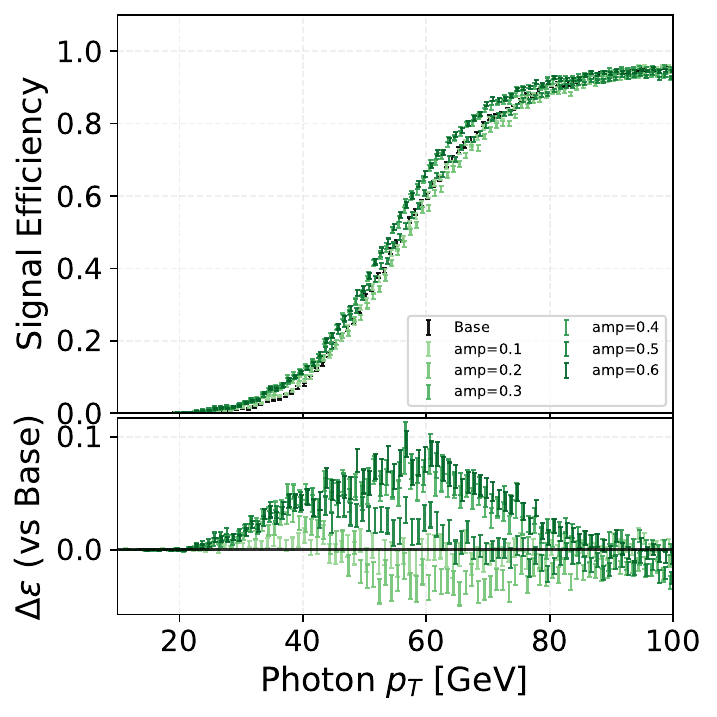} \\
(a) & (b)
\end{tabular}
\caption{\added{Soft scoring with exponential function.} (a) ROC curves showing baseline (black) and six amplitude variations (green gradient). (b) Turn-on curves at 0.1\% false positive rate with error bars and residuals vs.\ baseline. Amplitude 0.4 provides optimal performance ($\Delta E_{90} = -3.3$ GeV), with moderate gains across most amplitudes.}
\label{fig:soft_exp}
\end{figure}

\added{Figures~\ref{fig:soft_fd}, \ref{fig:soft_linear}, and \ref{fig:soft_exp} show ROC and turn-on curves for all configurations within each functional form. The turn-on curve residual panels demonstrate that performance gains are concentrated in the mid-$p_T$ regime (40--80 GeV), where the baseline ResNet already achieves moderate efficiency but soft scoring enables further improvement. At low $p_T$ ($<40$ GeV), where baseline efficiency is poor, soft scoring provides minimal benefit. At high $p_T$ ($>80$ GeV), the optimal amplitude for all configurations converge to slightly higher plateau efficiencies, confirming that optimized soft scoring allows improved efficiency without any degradation across the entire range.}

\added{The systematic evaluation demonstrates that soft scoring can provide meaningful performance improvements when properly configured. Higher amplitudes (0.4--0.6) consistently outperform lower values across all functional forms, yielding 3--5 GeV improvements in $E_{90}$ and 2--4\% gains in AUC. However, performance is sensitive to the amplitude parameter, with suboptimal choices potentially degrading results below baseline. Based on these results, we select \textbf{linear 0.5} as the optimal soft-scoring configuration for subsequent experiments.}

% ============================================================================
% SUBSECTION 4.3: AUXILIARY dR HEAD OPTIMIZATION
% ============================================================================
% ============================================================================
% SECTION 4.3: AUXILIARY dR HEAD OPTIMIZATION
% ============================================================================

\subsection{\newsection{Auxiliary $\Delta R$ Head Optimization}}
\label{subsec:dr_optimization}

\added{The auxiliary $\Delta R$ head introduces a multi-task learning framework where the network simultaneously predicts both class labels (signal vs.\ background) and the angular separation $\Delta R$ between photons in background events. This architecture tests two hypotheses: (1) explicit $\Delta R$ supervision may improve feature learning by encouraging the network to encode spatial shower structure, and (2) the predicted $\Delta R$ itself might serve as a useful discriminant. We systematically evaluate both possibilities through $\alpha$ ablation, where the combined loss is $\mathcal{L} = \alpha \mathcal{L}_{\mathrm{class}} + (1-\alpha) \mathcal{L}_{\Delta R}$.}

\added{All experiments use identical weight initialization (Section~\ref{subsec:experimental_protocol}) to ensure fair comparison. We test $\alpha \in \{0, 0.25, 0.5, 0.75, 1.0\}$, where $\alpha=0$ represents pure $\Delta R$ regression (no classification supervision) and $\alpha=1.0$ represents the baseline ResNet (no $\Delta R$ head). For each value, we evaluate two distinct scoring strategies: using the classification head output (primary analysis) and using the predicted $\Delta R$ as a discriminant (secondary analysis).}

\subsubsection{Classification Score Performance}

\added{Table~\ref{tab:dr_classification} presents results when using the classification head output as the discriminant across $\alpha \in \{0.25, 0.5, 0.75, 1.0\}$. We exclude $\alpha=0$ from this analysis as it trains only the $\Delta R$ regression head, leaving the classification head untrained.}

\begin{table}[htbp]
\newtab
\centering
\caption{\added{Auxiliary $\Delta R$ head performance using classification scores. $\alpha$ controls the balance between classification loss and $\Delta R$ regression loss. $\alpha=1.0$ is the baseline ResNet without $\Delta R$ head. Best configuration shown in bold.}}
\label{tab:dr_classification}
\small
\begin{tabular}{lcccc}
\hline
\textbf{$\alpha$} & \textbf{$E_{90}$ [GeV]} & \textbf{$\Delta E_{90}$ [GeV]} & \textbf{AUC [\%]} & \textbf{$\Delta$AUC [\%]} \\
\hline
0.25 & \textbf{74.1} & -5.2 & 49.2 & +3.7 \\
0.50 & 76.8 & -2.5 & 48.4 & +2.9 \\
0.75 & 74.5 & -4.8 & \textbf{49.4} & +3.9 \\
1.00 (Baseline) & 79.3 & --- & 45.5 & --- \\
\hline
\end{tabular}
\end{table}

\added{All configurations with the auxiliary $\Delta R$ head ($\alpha \leq 0.75$) outperform the baseline ResNet, demonstrating that multi-task learning with explicit $\Delta R$ supervision provides consistent benefits. Performance is relatively insensitive to the exact weighting: $\alpha=0.25$, $0.50$, and $0.75$ all yield 2--5 GeV improvements in $E_{90}$ and 3--4\% gains in AUC. The modest variation suggests that the auxiliary head acts primarily as a regularizer, encouraging the network to learn features relevant to spatial shower structure regardless of the precise loss balance.}

\begin{figure}[htbp]
\newfig
\centering
\begin{tabular}{cc}
\includegraphics[width=0.48\textwidth]{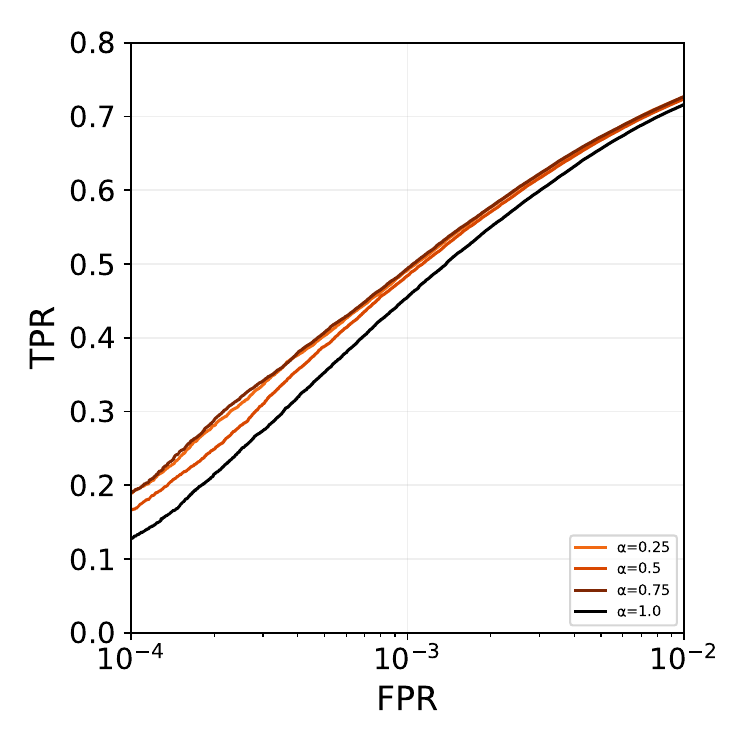} &
\includegraphics[width=0.48\textwidth]{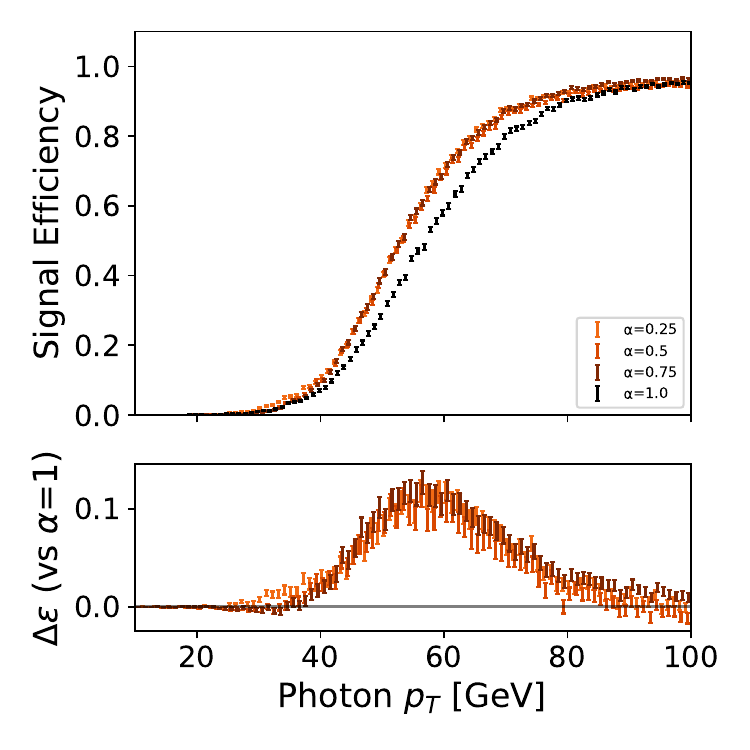} \\
(a) & (b)
\end{tabular}
\caption{\added{Auxiliary $\Delta R$ head performance using classification scores.} (a) ROC curves showing baseline (black, $\alpha=1.0$) and three $\alpha$ values (orange gradient). All configurations with the auxiliary head improve over baseline. (b) Turn-on curves at 0.1\% false positive rate with error bars and residuals vs.\ baseline. $\alpha=0.75$ (darkest orange) provides optimal balance of threshold performance and plateau efficiency.}
\label{fig:dr_classification}
\end{figure}

\added{We select $\alpha=0.75$ as the optimal configuration despite $\alpha=0.25$ achieving better $E_{90}$ (74.1 vs.\ 74.5 GeV). This choice is motivated by three factors: (1) $\alpha=0.75$ achieves superior AUC (49.4\% vs.\ 49.2\%), indicating better integrated performance across the full $p_T$ spectrum, (2) inspection of the turn-on curves (Figure~\ref{fig:dr_classification}) shows that $\alpha=0.75$ maintains higher efficiency in the plateau region ($p_T > 70$ GeV), which is more valuable for physics analyses, and (3) the higher classification weight ($75\%$ vs.\ $25\%$) provides a more interpretable architecture where classification remains the primary task.}

\subsubsection{Predicted $\Delta R$ as Classifier}

\added{To test whether the predicted $\Delta R$ itself can serve as a discriminant, we evaluate performance using $-\Delta R_{\mathrm{pred}}$ as the classifier score (lower predicted $\Delta R$ indicates higher signal probability). This analysis includes $\alpha=0$ (pure regression) but excludes $\alpha=1.0$ (no $\Delta R$ head).}

\begin{table}[htbp]
\newtab
\centering
\caption{\added{Performance using predicted $\Delta R$ as classifier. Score is $-\Delta R_{\mathrm{pred}}$ (lower $\Delta R$ = more signal-like). Pure regression ($\alpha=0$) trains only the $\Delta R$ head. Baseline shown for reference.}}
\label{tab:dr_prediction}
\small
\begin{tabular}{lcccc}
\hline
\textbf{$\alpha$} & \textbf{$E_{90}$ [GeV]} & \textbf{$\Delta E_{90}$ [GeV]} & \textbf{AUC [\%]} & \textbf{$\Delta$AUC [\%]} \\
\hline
0.00 (Pure Reg.) & 94.8 & +15.5 & 41.6 & -3.9 \\
0.25 & --- & --- & 32.3 & -13.2 \\
0.50 & --- & --- & 23.5 & -22.0 \\
0.75 & 86.4 & +7.1 & 45.0 & -0.5 \\
\hline
\multicolumn{5}{l}{\textit{Baseline (classification score, $\alpha=1.0$): $E_{90}=79.3$ GeV, AUC=45.5\%}} \\
\hline
\end{tabular}
\end{table}

\added{The results demonstrate that predicted $\Delta R$ performs poorly as a standalone classifier. Even pure $\Delta R$ regression ($\alpha=0$) substantially degrades performance relative to baseline ($E_{90}=94.8$ GeV vs.\ 79.3 GeV, $\Delta E_{90}=+15.5$ GeV). Configurations with joint training ($\alpha=0.25$, $0.5$) fail to reach 90\% efficiency within the studied $p_T$ range, indicating catastrophic failure of the $\Delta R$ predictor. Only $\alpha=0.75$, where classification dominates ($75\%$ weight), approaches baseline performance but still falls short ($E_{90}=86.4$ GeV, $\Delta E_{90}=+7.1$ GeV).}

\added{This analysis provides critical insight into the mechanism behind the auxiliary head's success: the benefit arises from regularization and improved feature learning, \textit{not} from $\Delta R$ encoding useful discriminative information. The network learns spatial shower structure through multi-task supervision, but this structure is most effectively exploited through the dedicated classification head rather than through direct $\Delta R$ comparison.}

\begin{figure}[htbp]
\newfig
\centering
\begin{tabular}{cc}
\includegraphics[width=0.48\textwidth]{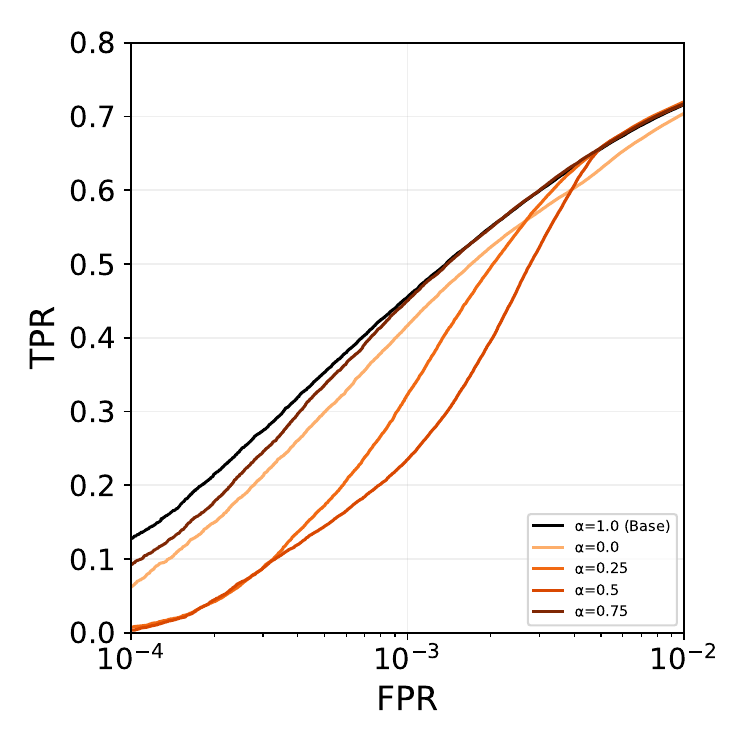} &
\includegraphics[width=0.48\textwidth]{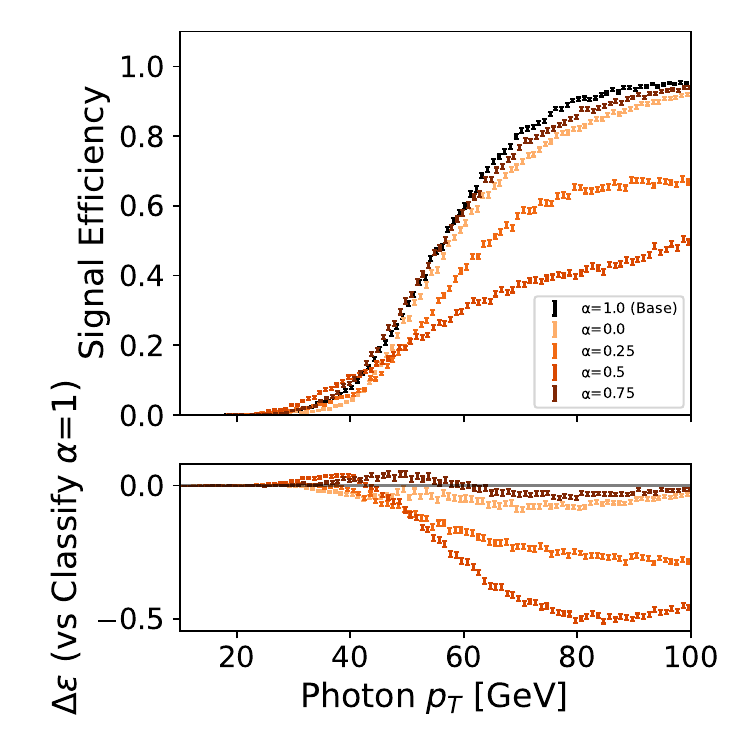} \\
(a) & (b)
\end{tabular}
\caption{\added{Performance using predicted $\Delta R$ as classifier.} (a) ROC curves for four $\alpha$ values (orange gradient). All perform substantially worse than baseline classification (Figure~\ref{fig:dr_classification}). (b) Turn-on curves showing poor discrimination, with $\alpha=0.25$ and $0.5$ failing to reach 90\% at all. Residuals computed relative to base line classification in black.}
\label{fig:dr_prediction}
\end{figure}

\added{Based on these results, we conclude that the auxiliary $\Delta R$ head is most effective as a regularization mechanism rather than as a source of additional discriminative features. We select $\alpha=0.75$ as the optimal configuration, achieving $E_{90}=74.5$ GeV (6.1\% improvement over baseline) while maintaining a clear primary focus on classification.}

% ============================================================================
% SUBSECTION 4.4: COMBINED METHODS
% ============================================================================
% ============================================================================
% SECTION 4.4: COMBINED METHODS
% ============================================================================

\subsection{\newsection{Combined Methods}}
\label{subsec:combined}

\added{Having established that both soft scoring (Section~\ref{subsec:soft_optimization}) and the auxiliary $\Delta R$ head (Section~\ref{subsec:dr_optimization}) independently improve performance, we now evaluate whether these techniques can be combined for further gains. The key question is whether the benefits are additive—yielding cumulative improvement—or whether the methods interfere, potentially degrading performance below that of individual approaches.}

\added{We compare four configurations using identical weight initialization (Section~\ref{subsec:experimental_protocol}): (1) baseline ResNet with no modifications, (2) soft scoring only (linear function, $s_{\mathrm{max}}=0.5$), (3) auxiliary $\Delta R$ head only ($\alpha=0.75$), and (4) both techniques combined. Table~\ref{tab:combined} presents results for all configurations.}

\begin{table}[htbp]
\newtab
\centering
\caption{\added{Performance comparison of combined methods. All configurations use identical initialization. Soft scoring uses linear function with $s_{\mathrm{max}}=0.5$. Auxiliary head uses $\alpha=0.75$. Best result per metric in bold.}}
\label{tab:combined}
\begin{tabular}{lcccc}
\hline
\textbf{Method} & \textbf{$E_{90}$ [GeV]} & \textbf{$\Delta E_{90}$ [GeV]} & \textbf{AUC [\%]} & \textbf{$\Delta$AUC [\%]} \\
\hline
Base           & 79.3 & ---   & 45.5 & ---  \\
+Soft          & \textbf{74.0} & -5.3 & 49.3 & +3.8 \\
+dR            & 74.5 & -4.8 & \textbf{49.4} & +3.9 \\
+Both          & 75.3 & -4.0 & 49.3 & +3.8 \\
\hline
\end{tabular}
\end{table}

\added{All three modified configurations substantially outperform the baseline, achieving 4--5 GeV improvements in $E_{90}$ and approximately 4\% gains in AUC. However, the combined approach (+Both) does not achieve fully additive benefits. While soft scoring alone yields $\Delta E_{90} = -5.3$ GeV and the auxiliary head alone achieves $\Delta E_{90} = -4.8$ GeV, their combination produces only $\Delta E_{90} = -4.0$ GeV—inferior to either individual method. The AUC metric shows similar behavior: +Both matches the individual methods (+3.8\%) rather than exceeding them.}

\begin{figure}[htbp]
\newfig
\centering
\begin{tabular}{cc}
\includegraphics[width=0.48\textwidth]{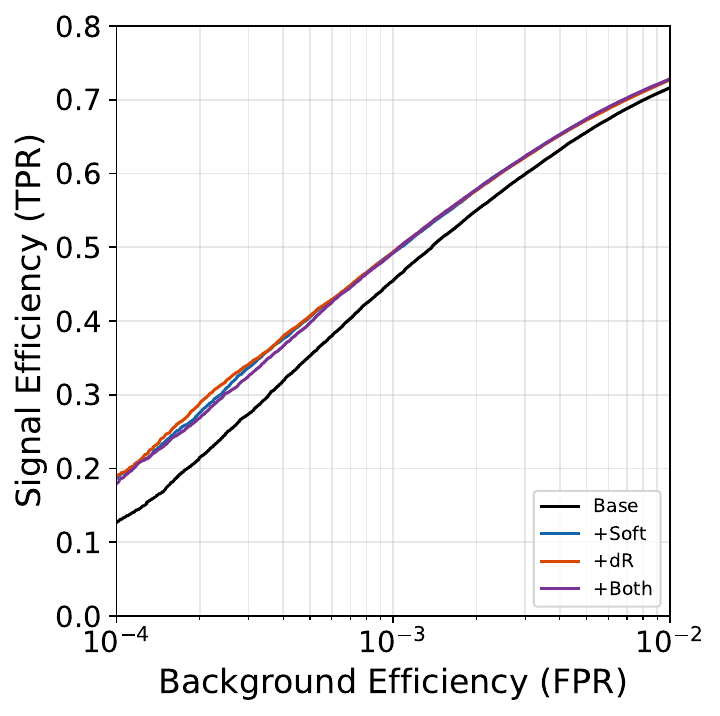} &
\includegraphics[width=0.48\textwidth]{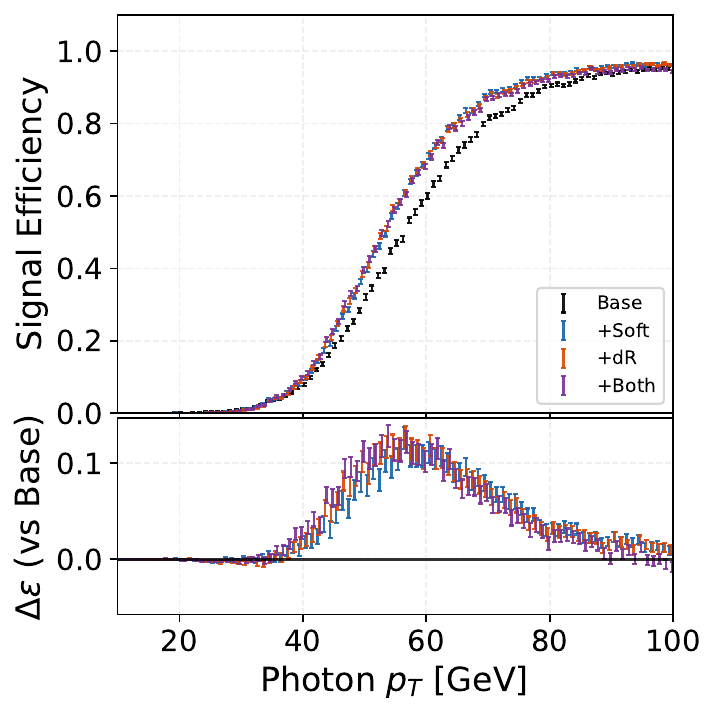} \\
(a) & (b)
\end{tabular}
\caption{\added{Combined methods comparison.} (a) ROC curves showing baseline (black) and three modifications: soft scoring (blue), auxiliary $\Delta R$ head (orange), and both combined (purple). All methods improve over baseline across the full FPR range. (b) Turn-on curves at 0.1\% false positive rate with error bars and residuals vs.\ baseline. Combined methods exhibit partial interference, performing between the two individual approaches rather than exceeding both.}
\label{fig:combined}
\end{figure}

\added{This result demonstrates partial destructive interference between the two techniques. Figure~\ref{fig:combined} illustrates this behavior through ROC and turn-on curves. Panel (b) shows that while all modifications improve efficiency across the mid-$p_T$ range (40--80 GeV), the combined method falls between the two individual approaches rather than surpassing both. The residual panel reveals that +Both achieves similar low-$p_T$ improvements to +Soft and +dR individually, but exhibits slightly reduced gains in the plateau region ($p_T > 70$ GeV).}

\added{The interference likely arises from conflicting optimization objectives. Soft scoring modifies the training targets to reflect label ambiguity, encouraging the network to assign intermediate scores to collimated backgrounds. The auxiliary $\Delta R$ head, meanwhile, trains the network to explicitly predict angular separation as an independent task. When combined, the network must simultaneously: (1) learn soft classification boundaries via modified labels, and (2) learn explicit $\Delta R$ prediction via multi-task supervision. These objectives may compete for network capacity in the shared convolutional layers, preventing either mechanism from fully expressing its individual benefit.}

\added{Despite the interference, the combined approach remains substantially better than baseline ($\Delta E_{90} = -4.0$ GeV, 5.0\% improvement), demonstrating that the techniques are not fundamentally incompatible—merely not fully synergistic. }

% ============================================================================
% SUBSECTION 4.5: ROBUSTNESS TO DETECTOR SYSTEMATICS
% ============================================================================
% ============================================================================
% SECTION 4.5: ROBUSTNESS TESTS
% ============================================================================

\subsection{\newsection{Robustness to Systematic Uncertainties}}
\label{subsec:robustness}

\added{To evaluate the stability of our optimized methods under realistic experimental conditions, we test robustness against two classes of systematic uncertainties: energy scale miscalibration and excess calorimeter noise. These tests assess whether performance improvements achieved under nominal conditions (Section~\ref{subsec:combined}) remain stable when the input data deviates from training conditions. We evaluate three configurations: baseline ResNet, soft scoring (linear function, $s_{\mathrm{max}}=0.5$), and auxiliary $\Delta R$ head ($\alpha=0.75$).}

\subsubsection{Energy Scale Uncertainties}

\added{Energy scale systematic uncertainties arise from imperfect detector calibration, affecting the absolute energy measurements across all calorimeter cells. We test six perturbation levels: $-10\%$, $-5\%$, $-2.5\%$, $+2.5\%$, $+5\%$, and $+10\%$, applied uniformly to all energy deposits. Models trained on nominal data are evaluated on scaled test sets without retraining, simulating the scenario where calibration drifts occur after deployment.}

\subsubsection{Calorimeter Noise}

\added{Excess calorimeter noise can arise from increased pileup conditions, degraded electronics, or environmental factors. The nominal training data already includes COCOA baseline noise (Section~\ref{sec:MC_Samples}) with layer-dependent Gaussian fluctuations. To test robustness beyond this baseline, we add additional Gaussian noise of $+10$ MeV and $+20$ MeV to all calorimeter cells. Given that the average energy per cell is approximately 300 MeV (with most cells empty or containing soft deposits), these perturbations represent substantial degradation: $+10$ MeV corresponds to $\sim$3\% relative noise, while $+20$ MeV represents $\sim$7\% relative noise—both significant increases over the existing COCOA baseline noise levels [13--75 MeV, layer-dependent].}

\subsubsection{Results}

\added{Table~\ref{tab:robustness} presents $E_{90}$ and AUC performance for all configurations under each systematic perturbation. Nominal results are included for reference.}

\begin{table}[htbp]
\newtab
\centering
\caption{\added{Robustness to systematic uncertainties. Models trained on nominal data (with COCOA baseline noise) are evaluated on perturbed test sets. Energy scale perturbations are applied uniformly to all cells. Additional Gaussian noise ($+10$ MeV, $+20$ MeV) is added beyond the existing COCOA baseline.}}
\label{tab:robustness}
\centering
\small
\begin{tabular}{lcccc}
\hline
\textbf{Perturbation} & \textbf{$E_{90}$ [GeV]} & \textbf{$\Delta E_{90}$ [GeV]} & \textbf{AUC [\%]} & \textbf{$\Delta$AUC [\%]} \\
\hline
\multicolumn{5}{c}{\textit{Baseline ResNet}} \\
\hline
Nominal & 79.3 & --- & 45.5 & --- \\
Scale $-10\%$ & 79.4 & +0.1 & 45.3 & -0.2 \\
Scale $-5\%$ & 79.3 & 0.0 & 45.4 & -0.1 \\
Scale $-2.5\%$ & 79.2 & -0.1 & 45.6 & +0.1 \\
Scale $+2.5\%$ & 79.8 & +0.5 & 45.4 & -0.1 \\
Scale $+5\%$ & 80.4 & +1.1 & 45.4 & -0.1 \\
Scale $+10\%$ & 84.2 & +4.9 & 45.2 & -0.3 \\
Noise $+10$ MeV & 83.6 & +4.3 & 45.0 & -0.5 \\
Noise $+20$ MeV & 89.4 & +10.1 & 42.6 & -2.9 \\
\hline
\multicolumn{5}{c}{\textit{Soft Scoring}} \\
\hline
Nominal & 74.0 & --- & 49.3 & --- \\
Scale $-10\%$ & 73.6 & -0.4 & 49.1 & -0.2 \\
Scale $-5\%$ & 73.7 & -0.3 & 49.2 & -0.1 \\
Scale $-2.5\%$ & 73.9 & -0.1 & 49.2 & -0.1 \\
Scale $+2.5\%$ & 74.0 & 0.0 & 49.4 & +0.1 \\
Scale $+5\%$ & 74.2 & +0.2 & 49.3 & 0.0 \\
Scale $+10\%$ & 75.0 & +1.0 & 49.3 & 0.0 \\
Noise $+10$ MeV & 74.6 & +0.6 & 49.2 & -0.1 \\
Noise $+20$ MeV & 78.5 & +4.5 & 47.9 & -1.4 \\
\hline
\multicolumn{5}{c}{\textit{Auxiliary $\Delta R$ Head}} \\
\hline
Nominal & 74.5 & --- & 49.4 & --- \\
Scale $-10\%$ & 74.3 & -0.2 & 49.2 & -0.2 \\
Scale $-5\%$ & 74.2 & -0.3 & 49.3 & -0.1 \\
Scale $-2.5\%$ & 74.2 & -0.3 & 49.4 & 0.0 \\
Scale $+2.5\%$ & 74.5 & 0.0 & 49.4 & 0.0 \\
Scale $+5\%$ & 74.7 & +0.2 & 49.4 & 0.0 \\
Scale $+10\%$ & 75.5 & +1.0 & 49.4 & 0.0 \\
Noise $+10$ MeV & 75.7 & +1.2 & 49.0 & -0.4 \\
Noise $+20$ MeV & 85.8 & +11.3 & 45.7 & -3.7 \\
\hline
\end{tabular}
\end{table}

\added{All configurations demonstrate excellent robustness to energy scale uncertainties within the $\pm 5\%$ range, with $E_{90}$ degradation limited to $\sim$1 GeV. Even at extreme $+10\%$ miscalibration, degradation remains modest ($\sim$5 GeV for Base, $\sim$1 GeV for optimized methods). This robustness arises from the network's focus on shower shape features and relative energy distributions rather than absolute energy scales. The soft scoring and auxiliary head methods maintain their performance advantages across all energy scale perturbations, demonstrating that the improvements are not artifacts of precise calibration.}

\added{Sensitivity to excess calorimeter noise is more pronounced. At $+10$ MeV additional noise, baseline performance degrades by 4.3 GeV, while soft scoring shows remarkable resilience with only 0.6 GeV degradation. The auxiliary head configuration exhibits intermediate sensitivity (+1.2 GeV). At $+20$ MeV, all methods degrade substantially, but soft scoring again demonstrates superior robustness: $\Delta E_{90} = +4.5$ GeV compared to $+10.1$ GeV (Base) and $+11.3$ GeV (+dR). This differential robustness suggests that soft scoring's continuous target representation may provide inherent regularization against input noise, while the auxiliary $\Delta R$ head's multi-task learning does not confer similar benefits.}

\added{The superior noise robustness of soft scoring is particularly valuable for experimental applications, where noise levels can vary with detector conditions, pileup, and environmental factors. The fact that soft scoring maintains substantial performance advantages even under $+20$ MeV excess noise ($E_{90} = 78.5$ GeV vs.\ 89.4 GeV for baseline) demonstrates that the method's benefits are not restricted to idealized conditions but persist under realistic detector degradation.}

\added{These robustness tests confirm that the performance improvements documented in Sections~\ref{subsec:soft_optimization}--\ref{subsec:combined} are stable against realistic systematic uncertainties. Both soft scoring and the auxiliary $\Delta R$ head maintain their advantages across energy scale variations and moderate noise increases, with soft scoring emerging as the most robust approach for challenging detector conditions.}
% ============================================================================
% SECTION 5: CONCLUSIONS
% ============================================================================

\section{Conclusions}

In this study, we generated and simulated photon and pion samples using the COCOA-HEP tool, a state-of-the-art detector simulation framework. We emulated the current baseline method for photon identification, a BDT classifier based on shower-shape variables, and introduced DNN and ResNet architectures as the natural next step toward improved performance.
Our results demonstrate that the ResNet significantly outperforms both the BDT and DNN across key metrics, particularly in the very low false-positive rate regime relevant for high-purity photon identification. \added{For example, ResNet achieves $E_{90} = 79.3$ GeV while neither baseline method reaches 90\% efficiency within the studied $p_T$ range.}
\changed{Furthermore, we showed that fine-tuning with soft scoring and incorporating angular separation information leads to additional gains, enabling the model to better discriminate the most challenging background cases without sacrificing signal efficiency.}
{We then systematically explored two approaches to push performance further: soft scoring (assigning continuous labels based on $\Delta R$) and an auxiliary $\Delta R$ regression head. Soft scoring with a linear function peaked at $E_{90} = 74.0$ GeV, while the auxiliary head with $\alpha=0.75$ achieved $E_{90} = 74.5$ GeV—both representing substantial 6\% gains over the ResNet baseline.}

\added{Interestingly, combining the methods proved counterproductive, yielding $E_{90} = 75.3$ GeV. This dip in performance relative to the individual models indicates a level of mutual interference, where the two strategies compete for the same network resources rather than complementing one another, and might require a simultaneous hyper-parameter scan for optimization. We also tested robustness under systematic uncertainties. Energy scale miscalibrations up to ±5\% caused minimal degradation (~1 GeV), but excess calorimeter noise proved more challenging. Adding 20 MeV Gaussian noise degraded the baseline by over 10 GeV, while soft scoring held up remarkably well with only 4.5 GeV degradation. The auxiliary head showed no such advantage, degrading by 11.3 GeV. This makes soft scoring particularly attractive for real detector conditions where noise levels fluctuate.}

These findings highlight the potential of modern deep learning architectures, combined with
\changed{physics-informed loss functions,}{physics-motivated training strategies,}
to improve photon identification in collider experiments.
Future work could extend these methods to full detector simulations and explore applications for real experimental data.
In contrast to previous deep-learning-based photon identification studies, which largely emphasized global shower properties or jet discrimination, our approach explicitly targets the substructure of overlapping electromagnetic showers in a symmetric calorimeter geometry. This method provides a clean, controlled benchmark for future photon identification developments at the High-Luminosity LHC and beyond.
\section{Acknowledgements}
The work of Y.F and L.B is supported by an ERC STGgrant (‘BoostDiscovery’, grant No.945878). We thank Dr. Rachel Jordan Hyneman and Dr. Luis Pascual Dominguez for their constructive feedback.
\appendix

\section*{Appendix}
\addcontentsline{toc}{section}{Appendix}  % Add to table of contents

\section{Model Architectures}
\label{app:architectures}

This appendix provides detailed architectural diagrams for the neural network models described in Section~\ref{sec:methods}.

\subsection{Dense Neural Network}
\label{app:dnn}
\begin{figure}[H]  % Use H to force placement HERE
\centering
\includegraphics[width=0.8\textwidth]{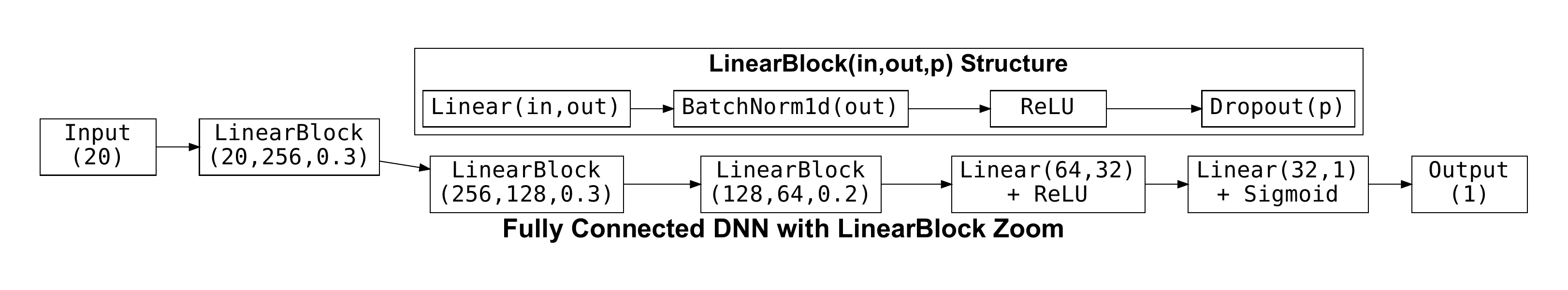}
\caption{Dense neural network architecture with LinearBlock components. Each LinearBlock consists of a linear layer, batch normalization, ReLU activation, and dropout with probability $p$.}
\label{fig:DNN}
\end{figure}

Figure~\ref{fig:DNN} shows the fully connected DNN architecture used as a baseline comparison to the BDT. The network consists of four fully connected layers with batch normalization, ReLU activations, and dropout regularization.

\subsection{ResNet Calorimeter Model}
\begin{figure}[H]  % Use H to force placement HERE
\centering
\includegraphics[width=0.9\textwidth]{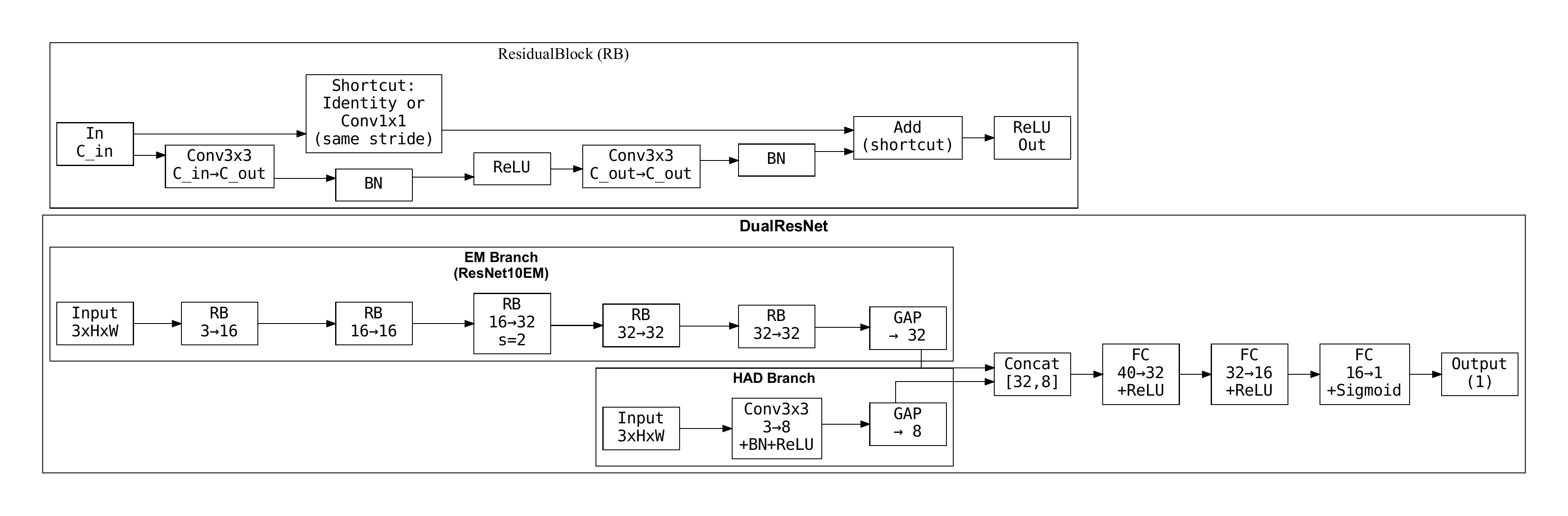}
\caption{DualResNet architecture with separate processing branches for EM and HAD calorimeter layers. Each branch uses residual blocks (RB) with skip connections. GAP = Global Average Pooling; BN = Batch Normalization.}
\label{fig:ResNet}
\end{figure}

Figure~\ref{fig:ResNet} illustrates the dual-branch ResNet architecture that processes electromagnetic (EM) and hadronic (HAD) calorimeter inputs separately before fusion.

\subsection{ResNet with Auxiliary Head}
\label{app:resnet_aux}
\begin{figure}[H]  % Use H to force placement HERE
\centering
\includegraphics[width=0.9\textwidth]{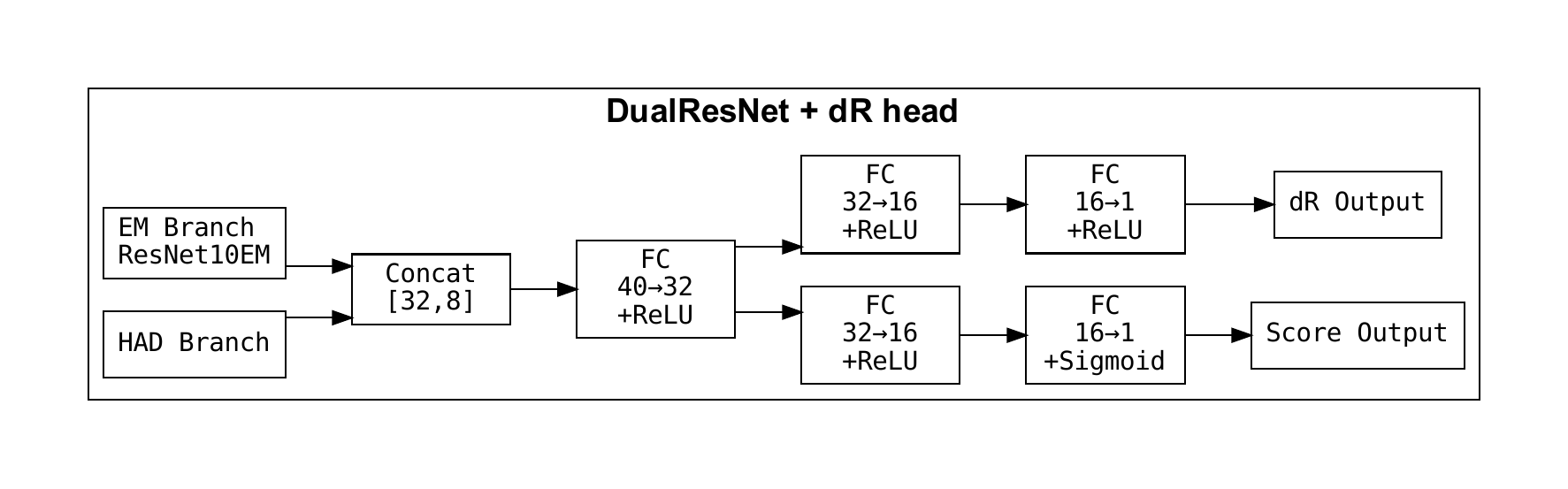}
\caption{DualResNet with auxiliary $\Delta R$ prediction head. The regression output provides additional supervision during training but is not used for final classification.}
\label{fig:ResNet_dR}
\end{figure}

Figure~\ref{fig:ResNet_dR} shows the multi-task extension with an auxiliary $\Delta R$ regression head that shares the convolutional backbone.

\clearpage  % Force page break before bibliography

\bibliographystyle{ieeetr}  % preserves numeric order, allows hyperlinks
\bibliography{sample}

\end{document}